\documentstyle[psfig,amssymb]{mn}

\newif\ifAMStwofonts

\begin{document}

\title[The Age and Metallicity of NGC~821
]
{
Spatially resolved stellar populations in the isolated elliptical NGC~821 
}
\author[
Proctor, Forbes, Forestell \& Gebhardt 
]
{
Robert N. Proctor$^{1}$, Duncan A. Forbes$^{1}$, Amy Forestell$^{2}$ \& Karl
Gebhardt$^{2}$ \\ 
$^1$ Centre for Astrophysics \& Supercomputing, Swinburne University,
Hawthorn VIC 3122, Australia\\
Email: rproctor@astro.swin.edu.au, dforbes@astro.swin.edu.au\\
$^2$ Department of Astronomy, University of Texas at Austin, C1400, Austin, TX~78712\\
Email: amydove@astro.as.utexas.edu, gebhardt@astro.as.utexas.edu}

\pagerange{\pageref{firstpage}--\pageref{lastpage}}
\def\LaTeX{L\kern-.36em\raise.3ex\hbox{a}\kern-.15em
    T\kern-.1667em\lower.7ex\hbox{E}\kern-.125emX}

\newtheorem{theorem}{Theorem}[section]

\label{firstpage}

\newpage

\maketitle

\begin{abstract}
We present the analysis of Lick absorption-line indices from three separate long-slit spectroscopic
observations of the nearby isolated elliptical galaxy NGC~821. The three data sets
present a consistent picture of the stellar population within one effective
radius, in which strong gradients are evident in both luminosity-weighted age and metallicity. The
central population exhibits a young age of $\sim4$~Gyr and a metallicity $\sim3$
times solar. At one effective radius the age has risen to $\sim12$~Gyr and
the metallicity fallen to less than $\sim\frac{1}{3}$ solar. 
The low metallicity population around one effective radius appears 
to have an exclusively red horizontal branch, with no significant
contribution from the blue horizontal branch evident in some globular 
clusters of the same age and metallicity.
Despite the strong central age gradient, we demonstrate that only a
small fraction ($\leq$10\%) of the galaxy's stellar mass can have been 
created in recent star formation events. We consider possible star formation
histories for NGC~821 and find that the most likely cause of the young
central population was a minor merger or tidal interaction that caused
NGC~821 to consume its own gas in a centrally concentrated burst of star
formation 1 to 4 Gyr ago.

\end{abstract}


\section{Introduction}
The advent of the Lick system of spectral indices (Worthey 1994) has provided
researchers with a useful tool in the disentanglement of age and metallicity
effects in the integrated spectra of stellar populations. This tool has
already been used to uncover interesting trends in the properties of the
stellar populations in the centres of galaxies (e.g. Worthey, Faber \&
Gonz\'{a}lez 
1992; Gorgas et al. 1997; Greggio 1997; Trager et al. 2000; Proctor \& Sansom 2002; 
Terlevich \& Forbes 2002; Proctor et al. 2004b). However, few studies
to date have probed significantly beyond the central regions. Most studies 
have hence only sampled a small, and perhaps biased, fraction of the galaxy mass.  
The few studies that probe off-centre regions (e.g. Mehlert et al. 2003;
Proctor 2002, S\'{a}nchez--Bl\'{a}zquez
et al. 2005) generally find metallicity gradients to be strong but both age and 
[Mg/Fe] gradients to be small or non-existent. Here we detail the application of 
Lick indices to the spatially resolved integrated spectra of NGC~821, which
we show to possess strong gradients in both metallicity \emph{and} age.

NGC~821 is a rare example of an isolated elliptical galaxy. At m--M =
31.91 (Tonry et al. 2001) it is the closest early-type galaxy in the
isolated galaxy sample of Reda et al. (2004). Reda et al. showed that NGC~821
conforms to the known scaling relations for elliptical galaxies (i.e. the Fundamental 
Plane and colour-magnitude relation). It has a moderate luminosity
(M$_V$ = --21.12) and highly elongated isophotes (E6). An isophotal
analysis indicates the presence of an edge-on stellar disk along the major axis (e.g.
Goudfrooij et al. 1994; Scorza \&
Bender 1995; de Souza, Gadotti \& dos Anjos 2004). It also 
reveals disk-like kinematics (Emsellem et al. 2004). Otherwise the galaxy 
shows no morphological fine structure (Michard \& Prugniel 2004;
Goudfrooij et al. 1994). In a recent catalogue of galaxy ISM properties, 
Bettoni, Galletta \& Garca-Burillo (2003) measured a dust mass that is at the very low end of the 
distribution for elliptical galaxies (Forbes 1991), while HI and molecular
hydrogen are undetected. 
The ISM properties of NGC 821 are therefore consistent with stellar
mass loss alone and do not require any external origin.

In previous studies, the luminosity-weighted age of the stellar population 
in the central region of NGC~821 has been measured to be  4.2~Gyr  (Denicolo 
et al. 2005), 7.7~Gyr (Trager et al. 2000), 11.5~Gyr (Caldwell, Rose \& Concannon 
2003) and 12.5~Gyr (Vazdekis, Trujillo \& Yamada 2004). Thus considerable
variation exists in the literature from claims of very young to very old ages.

In this work we use Lick indices to determine the radial profiles in
age, metallicity and `$\alpha$'--element abundance ratios in NGC~821 out to
one effective radius (1$r_{\rm e}$; the radius within which half the galaxy
light is contained). We use the $\chi^2$-minimisation technique of comparing a large number of
indices to SSP models to derive ages and metallicities as outlined in Proctor \&
Sansom (2002). We find a natural explanation for the range of ages measured in
previous studies. We also briefly discuss possible evolutionary histories for NGC~821.

\section{Data}
\label{data}
In this paper long-slit spectroscopy of three separate 
studies of NGC~821 are analysed. Each data set covered a wavelength range
suitable for the measurement of Lick indices. Trager et al. (1998) index 
definitions are used throughout.

\begin{figure}
\centerline{\psfig{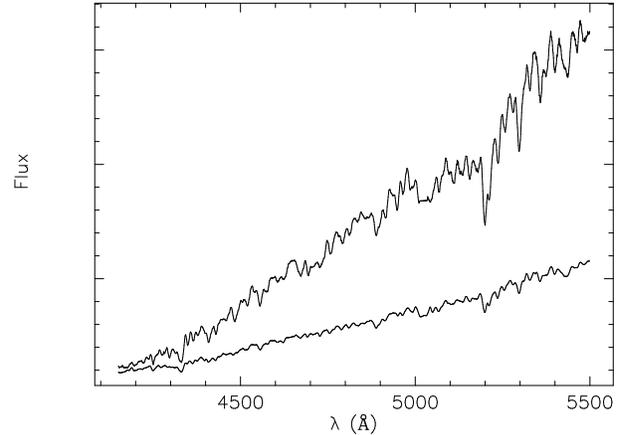}}
\caption{Gemini spectra from the central (upper) and outer (lower) regions
of NGC~821 are shown. The spectra have signal-to-noise of 80 and 20
respectively. The presented spectra have been broadened to the Lick
resolution of the H$\beta$ index (8.5~\AA) by convolution with a Gaussian.
An interpolation across a chip gap is evident at $\sim$5000~\AA.}
\label{spec}
\end{figure}

The first data set was obtained using GMOS on Gemini (North). 
These data were collected using a 0.5 arcsec slit positioned along the major
axis of the galaxy. The observations covered the wavelength range
4100~--~6930~\AA\  at 3.5~\AA\  resolution and a dispersion of ~1.77~\AA.
Three 30 minute exposures were obtained. The data were reduced using the 
Gemini pipe-line. Residuals to the wavelength calibration solution were $<$0.1~\AA. 
The three exposures were co-added prior to the extraction of spectra.
During extraction, the spectra were spatially binned to provide a signal-to-noise $>$20 at all
radial positions. Spectra from the central and outer regions are shown in Fig.
\ref{spec}. Reliable index estimates were obtained for 18 Lick indices from Ca4227 to
Fe5782. The GMOS observations were obtained as part of the author's
program to measure indices out to large radii ($\sim$2$r_{\rm e}$). 
Unfortunately, only 1.5 of the 8 hours requested exposure time was
actually observed.  This constrains the
index analysis to within $\sim$~1$r_{\rm e}$. Neither Lick-index nor
flux-calibration standard stars were observed.
The spectra could therefore not be flux calibrated by the normal means, and
the indices could not be fully calibrated to the Lick system. 
The effects of these uncertainties will be discussed in Section
\ref{calibrations}.

The second data set was obtained using the Hobby-Eberly Telescope (HET).
These data were originally collected for the purposes of
analysis of the spatially resolved stellar kinematics (Forestall \& Gebhardt 2005).
The data include long-slit spectroscopy along both major and minor axes using
a 1 arcsec slit. The observations covered the wavelength range 
4300~--~7250~\AA\ at 4.5~\AA\ resolution and a dispersion of ~1.95~\AA.
This wavelength range allowed the accurate measurement of 15 Lick indices from 
Fe4383 to TiO$_1$ (NaD was excluded due to the effects of interstellar
absorption).  The data were reduced and binned as described in Forestall \&
Gebhardt (2005), but were not flux calibrated. A signal-to-noise suitable for index
determination was achieved out to approximately 1$r_{\rm e}$.
As with the Gemini data, Lick index standard stars were 
not observed so the indices could not be fully calibrated to the
Lick system (see Section \ref{calibrations}). 

The final data set was obtained from the study of S\'{a}nchez--Bl\'{a}zquez
et al. (2005; hereafter SB05) as line index measurements.
The spectrum from the central region was also supplied for the 
purposes of calibration of the other two data sets  
(see Section \ref{calibrations}).
The data were collected using the 3.5~m telescope at the German-Spanish
Astronomical Observatory at Calar Alto. 
Long-slit spectroscopy was carried out with the 2.1 arcsec slit aligned to the parallactic 
angle, i.e. along a line at 25$^{\circ}$ to the 
major axis. The observations covered the wavelength range
3570~--~5770~\AA\ at 2.6~\AA\ resolution and a dispersion of 1.08~\AA.
This allowed accurate measurement of 19 Lick indices from H$\delta_A$ to Fe5335.
Although these data were of significantly lower signal-to-noise than the
Gemini and HET data, the spectra were accurately flux calibrated prior to
index measurements and fully calibrated to the Lick system.\\

\subsection{Calibrations}
\label{calibrations}
\begin{figure}
\centerline{\psfig{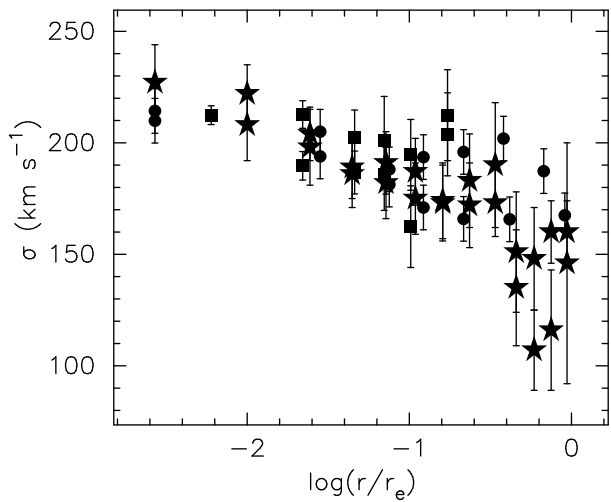}}
\caption{Radial profiles of velocity dispersion from all three data sets.
An effective radius of 50 arcsec (de Vaucouleurs et al. 1991) is assumed
throughout this work.
Gemini data are shown as filled stars, the HET data as filled circles and the
SB05 data as filled squares. The agreement between studies is generally
good.}
\label{kinematics}
\end{figure}

The calibration of Lick indices is a two stage process. The first is to 
compensate for the effects of internal velocity dispersion broadening. For the
Gemini data velocity dispersions were estimated by cross-correlation of the
galaxy spectra with stellar spectra observed with the same instrument 
configuration. The IRAF command \emph{fxcor} was used. For the HET data set 
measurements of kinematics were taken from Forestall \& Gebhardt (2005). For 
both Gemini and HET data sets indices were measured and velocity dispersion 
corrected to the appropriate resolution as described in Proctor \& Sansom (2002). For the SB05 
data set, both velocity-dispersion corrected indices and kinematics were 
supplied by the original authors. A comparison of the radially resolved 
velocity dispersions from the three studies is shown in Fig. \ref{kinematics}. 
Data points are shown at their projected distance from the galaxy centre in
this and all other radial plots in this work. Despite the differing slit widths and
orientations between the three studies, the comparison in Fig.
\ref{kinematics} can be seen to be 
generally good. Agreement is also good with the central velocity dispersions 
quoted in the literature (e.g. Bender, Saglia \& Gerhard 1994; 208 km
s$^{-1}$, Di Nella et al. 1995; 212 km s$^{-1}$, Denicol\`{o} et al. 2004; 212 km s$^{-1}$).

The second stage of the calibration to the Lick system is to measure offsets
between the indices observed in a number of Lick library stars and their 
published values. This process primarily compensates for differences
in flux-calibration between the observations and the stars used to construct
the SSP models. The  SB05 data set was fully
calibrated in this way (using 20 standard stars) and the average offsets
supplied with the indices. As noted above, the  SB05 spectra were also 
accurately flux-calibrated. The central spectrum
supplied with this data set was
therefore used to estimate flux-calibration curves for both Gemini and HET
data sets. This was achieved by third-order continuum fitting of the ratio of the fluxed
central spectrum from SB05 and the un-fluxed
central Gemini and HET spectra. The flux-calibration curves generated were
then applied to their respective data sets at all radii.
Indices were measured in both flux-calibrated and un-flux-calibrated
spectra for the purposes of comparison (see Section \ref{estimates}). We
shall refer to these as the fluxed and un-fluxed indices in the Gemini and
HET data sets. However, it should be noted the the SB05 spectra did not cover the wavelength range of indices redder than
Fe5335. The un-fluxed values of these indices were therefore used throughout 
this work. Average errors and the index offsets caused by the application of the flux
calibration curve are given in Table \ref{cals}. Both flux and Lick offsets are 
generally smaller than the average index errors. 

Since our analysis probes the low surface brightness outer regions of the
galaxy we must consider the effects of scattered light on our index
determinations. For the Gemini data this can be achieved by consideration of
count rates in regions of the slit that
are masked off (there is one region at each end of the slit and 
two approximately 
one-third and two-thirds along the length of the slit). Modelling these
count rates as scattered light with a Gaussian distribution, it is possible
to show that no more than 10\% of the light in the most affected regions
(the outermost points) can be from scattered light. Such contamination
levels would result in the systematic underestimation of the indices by an
amount of up to 10\% in the outer regions. However, since this is an
\emph{upper limit} on the contamination levels no attempt has been made to
correct indices for this affect. The impact of this uncertainty on our
derived ages and metallicities will be discussed in Section \ref{results}.

\subsection{Radial profiles in indices}
\begin{figure}
\centerline{\psfig{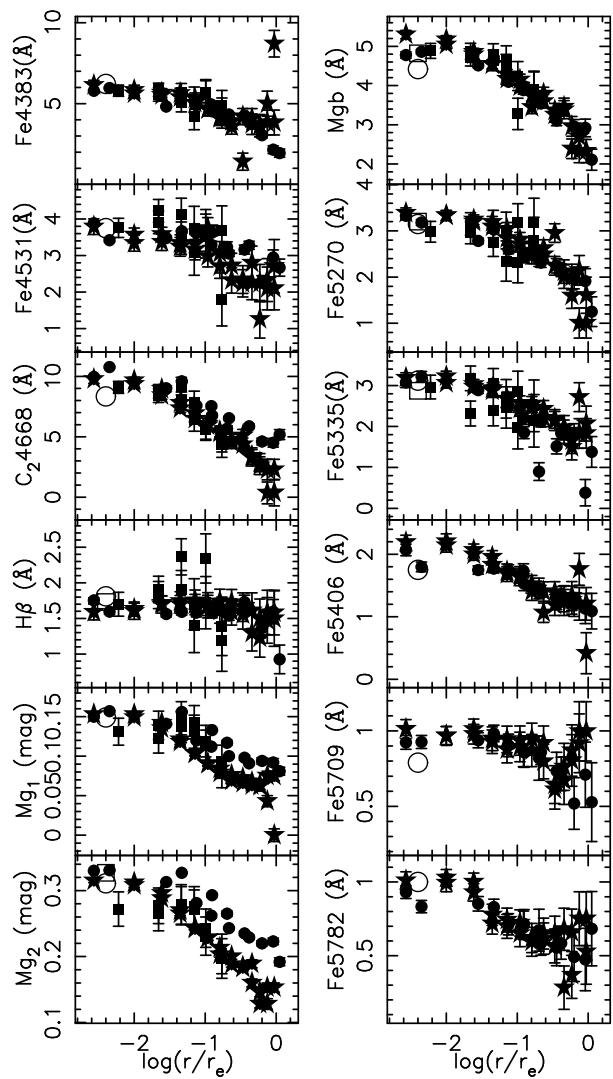}}
\caption{Radial profiles of twelve Lick indices are shown. 
Filled symbols are as Fig. \ref{kinematics}. The central values of Denicolo et
al. (2004) (open circles) and Gonz\'{a}lez (1993) (open squares) are also shown. 
Agreement between studies is generally good, with the exception of Mg$_1$ and Mg$_2$ (see text). 
Gradients are evident in all metallicity sensitive indices, while H$\beta$ shows no gradient.}
\label{radial}
\end{figure}

Radial profiles of twelve indices (which have been flux-calibrated but not corrected 
to the Lick system) are shown in Fig. \ref{radial}. 
Agreement between the three studies is generally good, despite the
differing slit widths and orientations.
Mg$_1$ and Mg$_2$ are notable exceptions to this good agreement (note the data shown
are the values from the flux calibrated spectra). The cause of this
discrepancy is not known, but is probably associated with poor
flux-calibration of the Gemini and HET data sets.
However, this problem does not effect our overall conclusions as the method
employed permits the exclusion of such troublesome indices (see Section
\ref{estimates}).
Negative gradients are evident in all metallicity sensitive 
indices. The H$\beta$ index, on the other hand, shows no significant gradient.

\begin{figure}
\centerline{\psfig{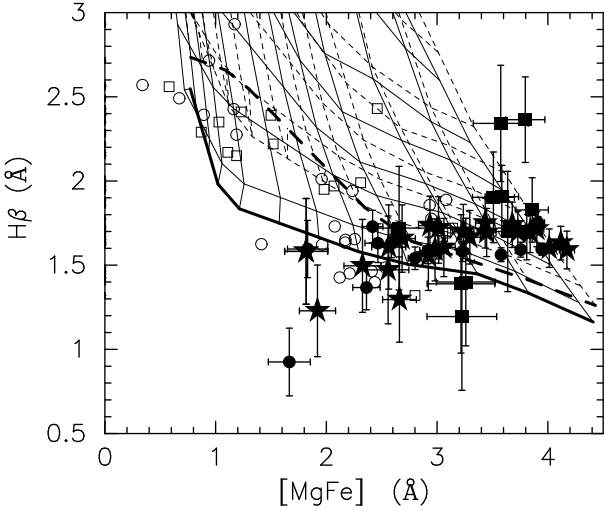}}
\caption{Plot of H$\beta$ against [MgFe]. Open circles are the Galactic
globular clusters of Proctor, Forbes \& Beasley (2004a). Open squares are
extra-galactic sub-populations from the meta-study of Strader et al. (2005).
Remaining symbols are as Fig. \ref{kinematics}. Grid lines show the SSP models of TMB03 (solid lines) and
TMK04 (dashed lines) for an `$\alpha$'--element enhancement of +0.3 dex.
The near horizontal lines show ages; from bottom to
top of 15, 12, 10, 8, 5, 3, 2, 1 Gyr. The thick lines denote the oldest age
grid lines. The near vertical lines show
metallicity; from right to left of +0.4, +0.25, 0.0 -0.25, -0.5,
-0.75, -1.0, -1.25, -1.5, -1.75, -2.0 and -2.25 dex. 
TMB03 models are those for populations with an exclusively red horizontal
branch. The poor agreement of the TMK04 SSP models (which include a blue
horizontal branch component) with the low metallicity population in NGC~821
is clearly evident. Both sets of models suggest that radial gradients are
present in both age and metallicity.}
\label{grids1}
\end{figure}

\begin{figure}
\centerline{\psfig{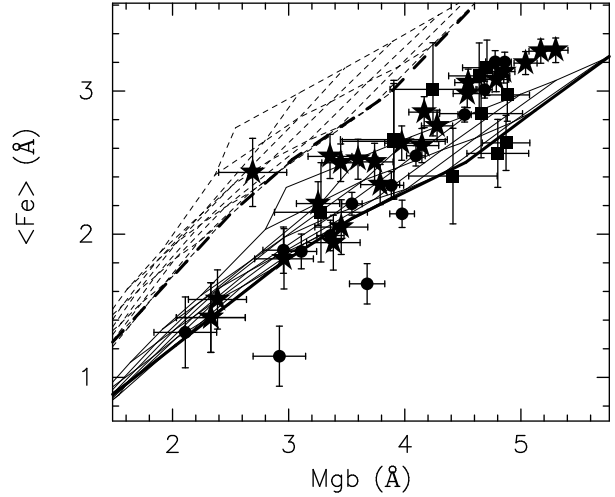}}
\caption{$<$Fe$>$ is plotted against Mgb. 
Symbols are as Fig. \ref{kinematics}. Grid lines show TMB03 models of solar
abundance ratio (dashed lines) and `$\alpha$'--element enhancement of +0.3 (solid lines).
Ages and metallicities of grid lines are as Fig. \ref{grids1}. The data
are generally consistent with grid lines with `$\alpha$'--element
enhancement of +0.3 dex.}
\label{grids2}
\end{figure}

Instructive index-index plots are shown in Figs \ref{grids1} and
\ref{grids2}\footnote{$[MgFe]=\sqrt{Mgb\times(Fe5270+Fe5335)/2}$ \\$<Fe>=(Fe5270+Fe5335)/2$}.
Fig. \ref{grids1} shows that in comparison to both Thomas, Maraston \&
Bender (2003; hereafter TMB03) and Thomas, Maraston \& Korn (2004; hereafter
TMK04) models, the galaxy appears to possess both age and metallicity gradients. 
This figure also demonstrates
that the central indices in this galaxy exceed those modelled by TMB03 or
TMK04. It was therefore necessary to linearly extrapolate
the models to an overall metallicity ([Z/H]) of +0.8 dex in order to fit
these data. It is interesting to note that the low metallicity data points
(with [MgFe]$<$2.0~\AA) show much better agreement with the red horizontal
branch models of TMB03 (solid lines in Fig. \ref{grids1}) than the models of
TMK04 (dashed lines) which include a blue horizontal branch component.
Indeed, the galaxy data points appear to follow the locus of those Galactic
globular clusters which possess exclusively red horizontal branch
morphologies (Proctor et al. 2004a).
Fig. \ref{grids2} also shows the consistency of the data with the SSP models
of TMB03 with `$\alpha$'--element enhancements of +0.3 dex, indicating that
NGC~821 possesses a similar abundance ratio to other bright elliptical galaxies
(e.g. Trager et al. 2000; Proctor \& Sansom 2002; Mehlert et al. 2003;
Proctor et al. 2004b) In the following section we outline the procedure 
by which values of age and [Fe/H] and the `$\alpha$'--element abundance 
ratio are determined.

\section{Determination of age and metallicity}
\label{estimates}

\begin{table*}
\footnotesize
\begin{center}
\begin{tabular}{lcccccc}
\hline
Index        & \multicolumn{2}{c}{Gemini}& \multicolumn{2}{c}{HET}
&\multicolumn{2}{c}{SB05}\\
             & Avg error & Flux offset   & Avg error & Flux offset & Avg error & Lick Offsets\\
\hline
H$\delta_A$  &  -     & -    &-    & -    &0.687&0.000   \\
H$\delta_F$  &  -     & -    &-    & -    &0.459&0.000   \\
CN$_1$       &  -     & -    &-    & -    &0.018&-0.013  \\
CN$_2$       &  -     & -    &-    & -    &0.021&-0.014  \\
Ca4227       &  0.204 & 0.004&-    & -    &0.304&-0.193  \\
G4300        &  0.354 & 0.108&-    & -    &0.509&-0.346  \\
H$\gamma_A$  &  0.402 & 0.513&-    & -    &0.607&0.568   \\
H$\gamma_F$  &  0.251 & 0.164&-    & -    &0.371&0.451   \\
Fe4383       &  0.492 & 0.082&0.125&-0.059&0.674&0.000   \\
Ca4455       &  0.254 & 0.019&0.068&-0.015&0.354&0.201   \\
Fe4531       &  0.367 & 0.006&0.110&-0.005&0.507&0.000   \\
C4668        &  0.531 &-0.013&0.190& 0.095&0.818&-0.682 \\
H$\beta$     &  0.197 &-0.017&0.089& 0.032&0.305&-0.104  \\
Fe5015       &  0.409 &-0.029&0.222& 0.275&0.665&0.000   \\
Mg$_1$       &  0.004 &-0.003&0.002& 0.027&0.019&0.013   \\
Mg$_2$       &  0.004 &-0.004&0.002& 0.016&0.028&0.002   \\
Mgb          &  0.183 &-0.001&0.122&-0.010&0.322&-0.157  \\
Fe5270       &  0.198 & 0.006&0.143&-0.067&0.365&0.000   \\
Fe5335       &  0.222 & 0.001&0.174&-0.047&0.445&0.000   \\
Fe5406       &  0.166 & 0.000&0.132&-0.022&-&-   \\
Fe5709       &  0.121 & -    &0.113&-     &-    &-\\
Fe5782       &  0.112 & -    &0.109&-     &-    &-\\
TiO$_1$      &  -     & -    &0.003&-     &-    &-\\
\hline
\end{tabular}
\end{center}
\caption{Calibration details. For Gemini and HET data sets average errors
and the difference in indices caused by the flux-calibration procedure
(given as un-fluxed minus fluxed) are shown. Flux offsets are generally
smaller than the average errors. For the SB05
data set average errors and the offset to the Lick system derived from
observations of standard stars are shown. Lick offsets are of similar
magnitude to the average errors.} 
\label{cals}
\normalsize
\end{table*}

For all three data sets measurements of log(age), [Fe/H] and
`$\alpha$'--abundance ratio (the derived parameters) were
made using the $\chi^2$--minimisation technique detailed in Proctor \&
Sansom (2002) and Proctor et al. (2004a,b). The `$\alpha$'--abundance ratio is parameterised by [E/Fe]; the
abundance ratio of a group of elements including C, O and Mg (see TMB03 for
details). Briefly, the technique for measuring the derived parameters involves the simultaneous
comparison of as many observed indices as possible to SSP models. The best fit is
found by minimising the deviations between observations and models in terms
of the observational errors i.e. $\chi$.
The rationale behind this approach is that, while all
indices show some degeneracy with respect to log(age), [Fe/H]
and [E/Fe], each index does contain \emph{some} information regarding
each parameter. In addition, such an approach should be relatively robust
with respect to many problems which are commonly experienced in the
analysis of spectral indices. These include poor flux calibration or sky
subtraction, stray cosmic rays and central emission (common in LINERs and AGN).
The method is also more robust with respect to the
uncertainties in the modelling of both the SSPs (e.g. the second parameter
effect in horizontal branch morphologies) and abundance ratio effects.
It was shown in Proctor et al. (2004a) that, at least for
Galactic globular clusters, the results so derived are significantly
more reliable than those based on only a few indices.\\

\subsection{Reliability of the method used}
In order to further demonstrate the improvement in reliability of derived
parameters achieved using the $\chi^2$--minimisation method over the more traditional use
of the H$\beta$--[MgFe] index combination, we report a simple, reproducible
test. First, we take the index values from a range of SSP models (here we use
metallicities from -2.25 to +0.5 dex in 0.25 dex steps and ages of 1, 1.5, 2,
3, 5 ,8, 12 and 15~Gyr). Then we perturb each index value of each model SSP to
represent the observational error. Here we take the error from a
randomly generated Gaussian distribution, the width of which
was taken from an actual SB05 observation. A low signal-to-noise observation was selected 
with Lick index errors close
to the maximum we deem suitable for age and metallicity determination (i.e.
H$\beta$ error $>$0.3~\AA). In
this way a realistic model data set is generated, whose true ages and
metallicities are well defined. Next, we estimate ages and metallicities using
the methods to be compared. 

\begin{figure}
\centerline{\psfig{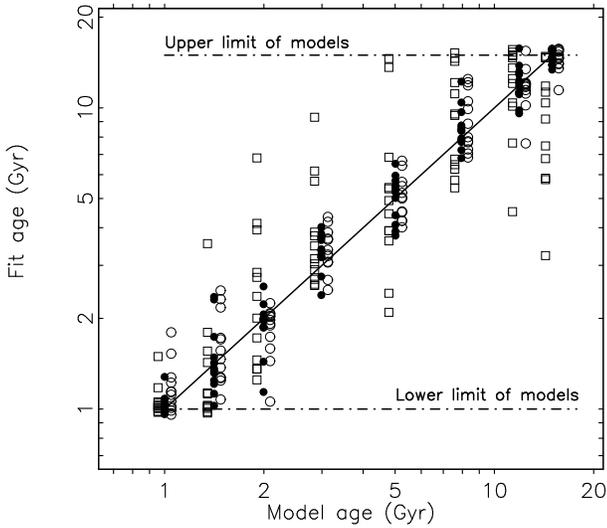}}
\caption{Comparison of ages derived using three different combinations of
indices. Filled circles represent results from  $\chi^2$--minimisation of
all 25 indices. These results  show a scatter of 0.07 dex about the model values.
Open circles represent results from  $\chi^2$--minimisation
of the remaining 20 indices when the five Balmer lines are excluded. The rms
scatter about model values in this case is 0.09 dex.
Open squares represent results from the H$\beta$--[MgFe] technique. These
results show an rms of 0.20 dex about model values.
The $\chi^2$--minimisation technique is clearly superior. Indeed, even when
all the Balmer lines are excluded from the fitting procedure, the results provide
a much improved accuracy over the H$\beta$--[MgFe] technique. }
\label{method}
\end{figure}

The results of such a comparison are shown
in Fig. \ref{method}. It is clear from Fig. \ref{method} that the $\chi^2$--fitting of
all 25 indices (with an rms scatter about model values of 0.07 dex or 17\%) 
is far superior in accuracy to values derived from the
H$\beta$--[MgFe] index combination (with an rms scatter about model values
of 0.20 dex or 58\%). Furthermore, the rms scatter about model values when
all the Balmer lines are omitted from the fitting procedure (0.09 dex or 23\%) is
also much smaller than that achieved using the H$\beta$--[MgFe] method.

The rms in metallicity about the model values was 0.1 dex both when all indices were included 
and when the Balmer lines omitted. 
For the H$\beta$--[MgFe] method this rms was 0.2 dex. For [E/Fe] rms scatters
were $\leq$0.1 dex both when all indices were included and when the Balmer
lines were omitted. [E/Fe] can not be estimated using the H$\beta$--[MgFe] method.

The improvement in accuracy of our $\chi^2$--minimisation technique
over the traditional H$\beta$--[MgFe] method
\emph{could} be understood simply in terms 
of the increased number of both age and metallicity sensitive indices employed. However,
we have also demonstrated that excellent results are still
obtained \emph{with all the Balmer lines excluded from the derivation}. This
emphasises the point above that even metallicity sensitive indices
contain \emph{some} information regarding age. This provides one of the main
motivations for our use of the $\chi^2$--minimisation technique.

In the analysis reported above, we have included all 25 Lick indices. Real
observations rarely include the full wavelength range, and often other
indices must also be omitted (see below). In addition, due to overlapping
band definitions, the errors of a number of indices are correlated.
We therefore tested more realistic
combinations of indices. These were based on the indices used in the
analysis of the three data sets below, and included only 15 indices
(i.e. excluding H$\delta_A$, H$\delta_F$, CN$_1$, CN$_2$, Ca4455, Mg$_1$, Mg$_2$, NaD
TiO$_1$ and TiO$_2$). It should be noted that the excluded indices include the
majority of those with correlated errors. The rms scatter when just these 15 indices were
included increased to 0.1 dex (26\% in derived age), still significantly
better than those derived by the H$\beta$--[MgFe] method.\\

\subsection{Model fitting}
In the following analysis we have compared the three sets of observational data
to two sets of SSP models; the red horizontal branch
(RHB) models of TMB03 and the models of TMK04.
As noted previously, Fig. \ref{grids1} shows that the TMK04 SSP models (which include a blue
horizontal branch component) are in poor agreement with the low metallicity 
population in NGC~821. The models of TMB03, which model purely red
horizontal branches, are in much better agreement. Given that other
differences between TMB03 and TMK04 models are small the results presented
in this work are those derived by comparison to the TMB03 (RHB) models. The
comparison to results derived using TMK04 models will be summarised in Section
\ref{results}.\\

\begin{figure*}
\centerline{\psfig{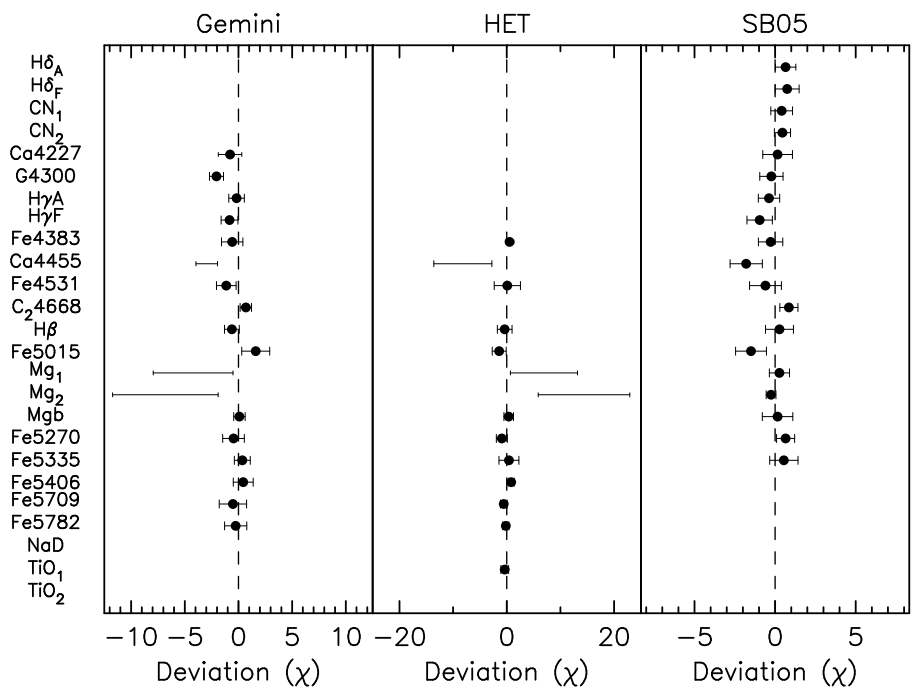}}
\caption{Deviations from best fit values for data compared to the models of
TMB03. Mg$_1$, Mg$_2$ and Ca4455 can be seen to be poor fits in the Gemini
and HET data. These indices were therefore excluded from the fitting procedure. 
Ca4455 is also depressed in the SB05 data but its exclusion does not affect the fitting results.}
\label{chis}
\end{figure*}

During the fitting procedure, a number of indices in each data set were found 
to suffer from problems that make them unsuitable for determining ages 
and metallicities (see Fig. \ref{chis}):

 - For both Gemini and HET data sets Mg$_1$ and Mg$_2$ were found to be
aberrant, even in the flux calibrated data (see Fig. \ref{radial}).

 - The Ca4455 index was found to be depressed in all three data sets.
 This effect has been noted by previous authors in other early-type galaxies
(e.g. Worthey 1992; Vazdekis et al. 1997; Trager et al. 1998;
Proctor \& Sansom 2002).

 - In some Gemini/GMOS spectra the Fe5015 index was found to be a poor fit
to the models. Subsequent inspection showed that this occurred when the index fell on a 
chip-gap interpolation (see Fig. \ref{spec}).

 - Some other individual indices were also found to be poor fits to the data.
These were generally associated with sky-line residuals and vignetting
(edge) effects in the data.\\

The indices identified above as giving poor fits to the models were excluded
from the fitting procedure. The average deviations of indices to the best
fit model values are shown in
Fig. \ref{chis}, where the Mg$_1$, Mg$_2$ and Ca4455 discrepancies are
clearly visible. 

The effects of flux-calibration and the application of the Lick offsets of 
SB05 to the Gemini and HET data sets were also investigated. The small
offsets in indices caused by flux-calibration (Table 1) had only small effects on the 
derived parameters. For the Gemini data a difference of +0.05$\pm$0.05 
dex was found in all three parameters. For the HET data there was an offset
in metallicity of $\sim$--0.1$\pm$0.05 dex when flux-calibration was applied.
Differences of +0.05$\pm$0.05 dex were found in log(age) and [E/Fe].
The $\chi^2$ values of the best fits were found to be marginally lower for
the flux-calibrated data. We therefore present data for flux-calibrated
spectra in the following. 

Differences caused by the application of Lick offsets of SB05 to the Gemini
and HET data sets (Table 1) also only caused small offsets in derived
parameters (log(age): --0.08$\pm$0.09, [Fe/H]: 0.00$\pm$0.07 and 
[E/Fe]: +0.05$\pm$0.05). The $\chi^2$ values of the best fits were
found to be marginally lower when the Lick offset was \emph{not} applied to
the Gemini and HET indices. These offsets were therefore not applied to
these data sets in the results that follow.

\section{Results}
\label{results}
\begin{figure*}
\centerline{\psfig{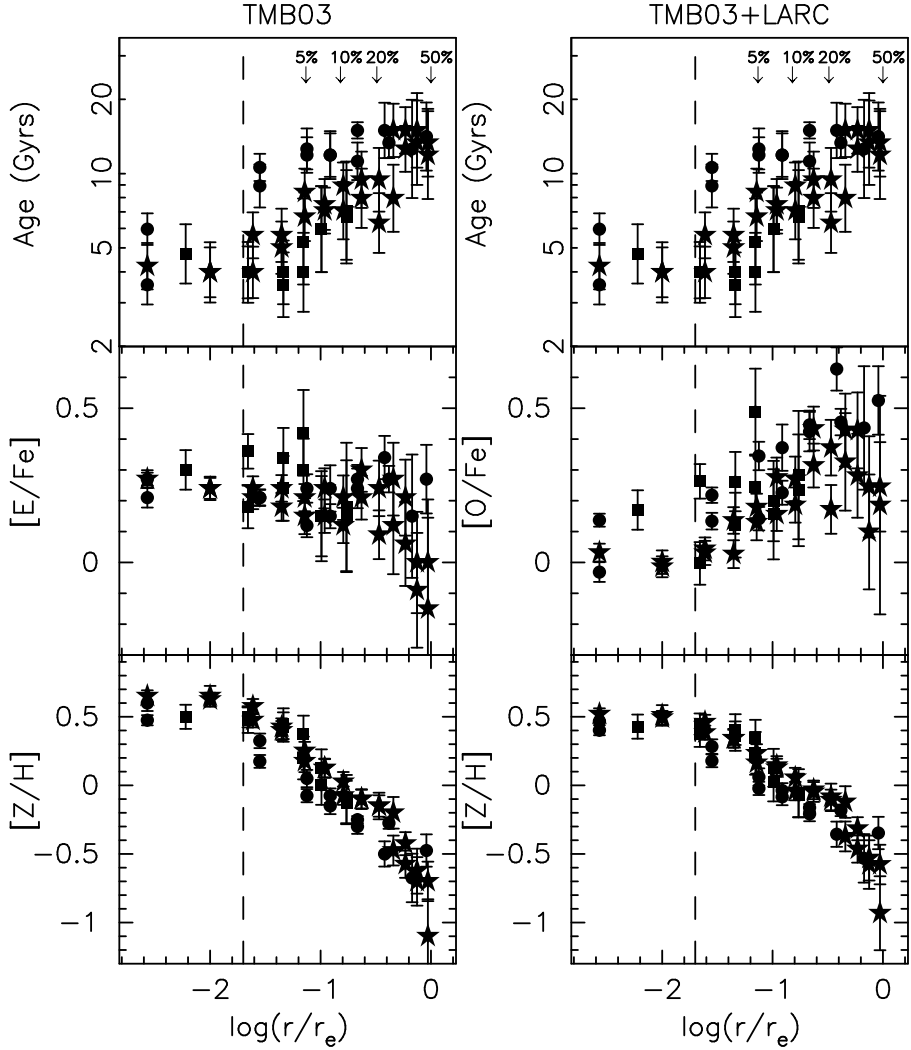}}
\caption{Radial profiles in derived parameters are shown for all three data
sets. Symbols are as Fig. \ref{kinematics}. The approximate extent of the
seeing discs of the three studies are identified by a dashed line.
The radii within which 5\%, 10\%, 20\% and 50\% of the galaxy's
stellar mass resides are also indicated. Age and metallicity values in the left-hand 
plots are those derived directly from TMB03 SSP models, while values in the 
right-hand plots have been LARC corrected (see text). Gradients are evident 
in both age and [Z/H]. However, no gradient is present in [E/Fe] (left-hand plot), 
while the corrected [O/Fe] (right-hand plot) shows a clear gradient.}
\label{agez}
\end{figure*}

The results of $\chi^2$--fitting of the data to the RHB models of TMB03 
are shown in the left-hand plots of Fig. \ref{agez}. The Gemini and HET data 
fitted were those that had been flux-calibrated but without the application of 
the Lick offsets of SB05. Errors are estimated using 50 Monte Carlo
realisations of each data point.

Fig. \ref{agez} shows that, generally, agreement 
between the three studies is good, despite the differing slit widths and 
orientations. Strong gradients in both age and [Z/H] are evident. 
Fig. \ref{agez} also shows that the age gradient extends from the edge of
the seeing affected central region out to log(r/$r_{\rm e}$)$\sim$0.5, with
the population outside this region being uniformly old (i.e. $\sim$12~Gyr).
The radial [Z/H] gradient (--0.79 dex/dex for points with --1.7$<$log(r/$r_{\rm e}$)$<$0.0) 
is extremely strong (e.g. Proctor 2002; Mehlert et al. 2003), suggesting 
that the young, central population formed from highly enriched gas.

Age and metallicity determinations were also made using the Gemini data with 
all the indices increased by 10\%. This was done in order to test the impact of 
scattered light on the derived parameters in the outer regions of the galaxy 
(Section \ref{calibrations}). The resultant ages for the outermost points were 
identical to those presented here, while [Fe/H] and [Z/H] were increased 
by 0.1 dex. Given the good agreement between the ages and metallicities 
from the Gemini data and those from the other two studies (for which we 
are not able to estimate scattered light directly), we conclude that the 
contamination is also $\leq$10\% in all three studies. The metallicity 
gradient given above may therefore be overestimated by as much as 
0.1~dex/dex due to scattered light contamination. In addition, the use of 
TMK04 models results in a slight ($\sim$~10\%) reduction in the 
metallicity gradient (as well as an increase in central age of
$\sim$2~Gyr to 6~Gyr). However, even when both scattered light and modelling
uncertainties are considered, the metallicity gradient remains steeper than
--0.5 dex, leaving the qualitative results outlined below unchanged.

Also indicated in Fig. \ref{agez} are estimates of the radii within which 5\%, 10\%, 20\%
and 50\% of the galaxy's stellar mass resides. These radii were estimated
using the luminosity profile of Ravindranath et al. (2001), an effective radius of 50
arcsec (de Vaucouleurs et al. 1991) and the assumption of a constant stellar mass-to-light ratio. It is
evident from Fig. \ref{agez} that no more than $\sim$10\% of the stellar
mass can have been involved in the star formation event responsible for the
young central age.

Fig. \ref{agez} also shows that the [E/Fe] values in NGC~821 are consistent 
with no gradient. However, Proctor et al. (2004b) expressed concerns
regarding the values of [Z/H] and [E/Fe] that are derived directly from the
SSP models. These concerns stem from consideration of the oxygen
abundance ratios observed in both the solar neighbourhood, the bulge and the
ISM of distant galaxies. 

\subsection{The local abundance ratio}
Oxygen is an important element for 
two reasons: It is by far the most abundant metal -- constituting $\sim$55\% 
of the total metal content of the Sun (Asplund, Grevesse \& Sauval 2004). It is a good tracer of Type II supernovae, since its local
Galactic abundance pattern is consistent with it having no other nucleo-synthetic 
sources  (Qian \& Wasserburg 2001). The oxygen abundance [O/Fe] in the solar 
immediate neighbourhood (the thin and thick discs and the halo) is approximately +0.6
dex at metallicities below [Fe/H]=--1. Between [Fe/H]=--1.0 and [Fe/H]=0.0
[O/Fe] falls from +0.6 dex to 0.0 dex (Chiappini, Romano \& Matteucci 2003). 
Bensby, Felzing \& Lundstr{\o}m (2003) show that this downward trend in [O/Fe] continues 
at [Fe/H]$>$0.0, while the abundance ratios of other $\alpha$--elements
(e.g. [Mg/Fe]) remain almost constant at the solar value. Fulbright, Rich \&
McWilliam 2004 have recently shown a similar trend in [O/Fe] with [Fe/H]
in Galactic bulge stars, but with [Mg/Fe] elevated with respect to disc stars. 
The low [O/Mg] ratios implied by these results have also been observed in
the ISM of early-type galaxies (Humphrey \& Buote 2005 and references therein).

These observations present a challenge to current models of Galaxy formation, and
indicate that new physics is required (e.g. see Fulbright et al. 2004).
Our concerns are therefore twofold:\\

1) The [Z/H] values assigned to the TMB03 (or any other) SSP models of indices
do not include oxygen. Consequently, using the up-to-date solar abundances
of Asplund et al. (2004), we calculate that the [Z/H] values at low
metallicities are understated by as much as 0.2 dex, while at high
metallicities they are overstated by as much as 0.15 dex.\\

2) It is a commonly held belief that \emph{all} [$\alpha$/Fe] ratios
([O/Fe], [Ne/Fe], [Mg/Fe], [Si/Fe] etc) can act as 
indicators of formation time-scale in stellar populations. 
Key assumptions in the reasoning for this belief is that the 
[Fe/H] of a population is a good proxy for its time of formation (due to
Type Ia supernova production of Fe), while
[$\alpha$/H] is a proxy for the total amount of star formation preceding
the formation of the population in question (due to the Type II supernova production of
$\alpha$--elements). The constant [Mg/Fe] and falling [O/Fe] with [Fe/H] in high 
metallicity disc and bulges stars of our Galaxy 
clearly indicates a breakdown in one or
more of these assumptions. We must therefore either abandon the
idea that the [Fe/H] of a population is a proxy for its time of
formation (and the belief that [$\alpha$/Fe] ratios can act as
indicators of formation time-scales), or we must accept that 
not \emph{all} [$\alpha$/Fe] ratios can act as
indicators of formation time-scale in stellar populations.



As a result of these considerations we have applied corrections to the
values of [Z/H] to include the effects of the oxygen abundance
variations in the stars used to construct the SSP models. Both corrected and
uncorrected values are presented in the results that follow.

The correction to
[Z/H] was carried out by applying an offset ($\Delta$Z) to the values derived 
directly from the TMB03 models using the up-to-date solar abundances of
Asplund et al. (2004), the [O/Fe] trend outlined previously. $\Delta$Z was given by:\\

$\Delta$Z=+0.175 (for [Fe/H]$<$--1.0)\\

$\Delta$Z=--0.175[Fe/H] (for --1.0$\leq$[Fe/H]$\leq$0.0)\\

$\Delta$Z=--0.33[Fe/H] (for [Fe/H]$>$0.0)\\

Rather than correct [E/Fe] for local trends in [O/Fe], we have chosen to
represent the `$\alpha$'--element abundance ratio \emph{by} [O/Fe]. Both
[E/Fe] and [O/Fe] are presented in the results that follow.
[O/Fe] was calculated using:\\

[O/Fe]=[E/Fe]+0.35 (for [Fe/H]$<$--1.0)\\

[O/Fe]=[E/Fe]--0.35[Fe/H] (for --1.0$\leq$[Fe/H]$\leq$0.0)\\

[O/Fe]=[E/Fe]--0.60[Fe/H] (for [Fe/H]$>$0.0)\\

It is important to note that the expressions given above assume that the [O/Fe] 
with [Fe/H] trend in the stellar population under consideration is the same as that 
in the solar-neighbourhood. This assumption is supported by the fact that in [O/Fe] 
with [Fe/H] trends of stars in the bulge of our own Galaxy show an identical
trend to the stars in the disc (Bensby et al. 2003; Fulbright et al. 2004), even 
though the [Mg/Fe] of the two populations differ
significantly. We shall refer to the correction to [Z/H] and the transform from 
[E/Fe] to [O/Fe]  as the `local abundance ratio correction' (LARC).\\

There are two significant effects of the local abundance ratio
correction on our results for NGC~821. The effect on the [Z/H] gradient is 
to reduce it somewhat (to $\sim$--0.60 dex/dex), although this is still an 
extremely strong 
metallicity gradient. The effect on the slope of the enhancement parameter
(now [O/Fe]) is quite profound, with a gradient evident in [O/Fe] that was
not present in [Mg/Fe]. The gradient is such that the
young, central population is significantly less enhanced in oxygen with
respect to iron than the older,
outer regions. This is, at least qualitatively, more in line with
expectations, given that we might expect the young, metal-rich population in
the centre to have formed from
gas that had been enriched by Type Ia supernovae over a longer time-scale than
the old population in the outer regions of the galaxy.\\

\begin{figure}
\centerline{\psfig{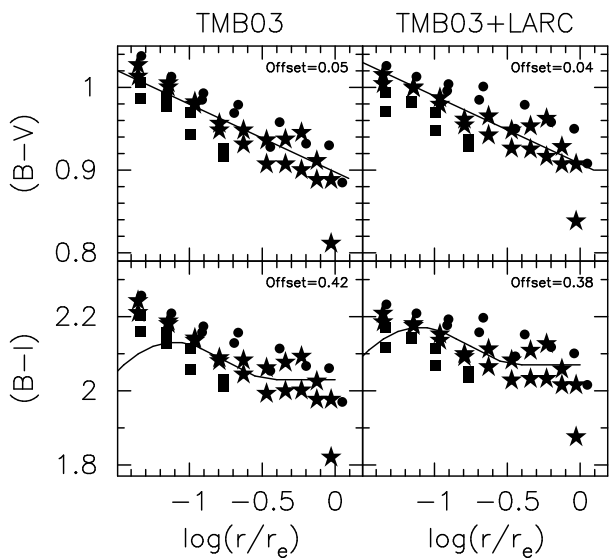}}
\caption{Colours predicted by age and metallicity from TMB03 SSP models and
the colours of TMK04. Symbols are as Fig. \ref{kinematics}.
The radial trends in colours from  Goudfrooij et al.
(1994) are also shown (lines). However, offsets (given top-right in each plot) have been
applied to the  Goudfrooij et al. data in order to match our results. These
are consistent with the effects of extinction but other contributory factors
may be present, such as modelling uncertainties in the SSP models.}
\label{colours}
\end{figure}

As a further test of the derived values of age and metallicity, colours were
estimated from the models of TMK04 
using the age and [Z/H] values shown in Fig. \ref{agez}. Colours for age and
[Z/H] values derived directly from the TMB03 SSP models are shown in the
left-hand plots of Fig. \ref{colours}, while colours from LARC
values are shown in the right-hand plots. These estimates are compared to the
photometric observations of Goudfrooij et al. (1994).
It was necessary to apply significant offsets (given in the plots) to the 
Goudfrooij et al. (1994) data in order to obtain a match between our estimates and the
observations. The sense of these offsets are consistent with
the effects of internal extinction, although other systematic effects (e.g. SSP modelling
errors) may be present. However, once this offset is applied the match between
observations and data are extremely good, i.e. we recover the relatively
shallow colour gradients of Goudfrooij et al. (1994) extremely well, despite
the strong age and metallicity gradients present in our spectroscopic results.

\section{Discussion}
The central luminosity-weighted age of $\sim$4~Gyr found in this work is consistent 
with the young ages found in
some previous studies. However, as noted in the introduction, other studies
have found \emph{old} ages. We find a natural explanation for these results when
we consider the strong age gradient and the varying aperture size between
studies. The results presented here for the central regions are for small
regions (0.5x0.5 arcsec, 1.0x0.3 arcsec and 2.1x1.1 arcsec for the Gemini, HET and 
SB05 data sets respectively). Denicolo et al.
(2004), who obtain a similarly young age of 4.2 Gyr, used an aperture 1.5x12.0 arcsec,
while the Trager et al. (2000) age of 7.7 Gyr was for a 2.0x12.0 arcsec. 
The two studies that found old ages also used large apertures -- Vazdekis et
al. (2004) quote an age of 12.5~Gyr within an aperture of 1.6x10 arcsec, while
Caldwell et al. (2003) find an age of 11.5~Gyr using a 3x5 arcsec slit. A
plot of effective aperture size (using the J{\o}rgensen, Franx \&
Kj{\ae}rgaard 1995 definition) the against the measured age shows a
correlation, albeit with significant scatter. We therefore conclude that the 
presence of the age gradient and the varying aperture shapes/sizes between the
various studies plays a part in causing the range of ages in the literature.\\

We next turn our attention to possible star formation histories in NGC~821.
The young central age clearly suggests relatively recent star formation,
although there is no evidence for \emph{current} star formation. It is
therefore appropriate to consider the source of the gas that fuelled this star formation.
By making some simple assumptions regarding the relative luminosities of
young and old populations we estimate that the mass involved in the more
recent burst to be
between a few times $\sim$10$^8$M$_{\odot}$ (assuming the burst to be 1~Gyr old) and a
few times 10$^{10}$M$_{\odot}$ (assuming the burst to be 4~Gyr old). We
therefore constrain the mass involved in the recent starburst to be between
$\sim$0.2\% and $\sim$10\% of the overall galactic mass.
These values above are in stark contrast to the current mass of dust
($\sim$2.5$\times$10$^5$M$_{\odot}$; Bettoni et al. 2003) and HI gas
($<$10$^8$M$_{\odot}$; Georgakakis et al. 2001). It is therefore clear that
the gas involved in the recent burst was either almost completely consumed
in star formation or was ejected from the galaxy as a wind.\\

Next, we consider a number of sources for the gas that fuelled the recent
starburst; a major merger, a minor gas-rich merger and in-situ
gas.

The first consideration is the extremely strong metallicity gradient present in NGC~821.
Such a strong gradient is inconsistent
with merger models which predict small gradients due to the `washing-out' of
the ambient gradients in the turbulent mixing of the merger (Kobayashi 2004). 
The gradient in NGC~821 ($\sim$--0.7 dex/dex) is therefore significantly greater than is achieved
in merger models. In addition, NGC~821 lacks the tidal tails, shells and
plumes expected to persist for $\sim$3~Gyr after a major merger (e.g.
NGC~1700; Brown et al. 2000). Thus, a recent major merger seems unlikely.

An alternative explanation is that of  the accretion
of a small, gaseous, satellite galaxy. However, in this scenario, the extremely high
central metallicities are hard to achieve since the small accreted galaxy
would, presumably, have contained \emph{low} metallicity gas. 

The final scenario we consider is that of a central burst of star-formation 
fuelled by in-situ gas from the galaxy itself.
This could have come from the smooth accretion of cooled halo gas onto a disc 
as predicted by hierarchical models (e.g. Meza et al. 2003). It is therefore  possible that the 
recent starburst was associated with gas from the embedded stellar disc 
(Goudfrooij et al. 1994; Scorza \& Bender 1995; de Souza et al. 2004). 
However, Scorza \& Bender (1995) show the disc to be highly inclined, with an 
inclination of $\sim$72$^{\circ}$. Therefore, since the young ages are seen to extend
to the same relatively large radii on both major and minor axes (Fig.
\ref{agez}), it is unlikely that the recent starburst was \emph{confined} to the disc.

Regardless of the source of the gas within the
galaxy, we would expect the younger population to have `$\alpha$'--element
abundances depressed with respect to the outer regions. 
In other words, we would expect the time-scale of star formation to be longer
in the young, high-metallicity, central region of NGC~821 than in the old,
low-metallicity outer regions.
This is clearly not the case for the [E/Fe] values derived directly from TMB03
models. It \emph{is} the case, however, for our estimates of [O/Fe]. 
We are therefore able to construct a recent star formation history for
NGC~821 that is consistent with both spectroscopically determined ages and
metallicities and the photometric data by
assuming that between 0.2\% and 10\% of the galaxy's mass was
formed 1~Gyr to 4~Gyr ago, respectively, from the galaxy's own gas. 
The paucity of gas remaining within the galaxy favours the lower end of this 
mass range and younger ages for the recent starburst. This, then, represents our favoured scenario.

\section{Conclusions}
We have presented the analysis of Lick indices from three separate long-slit
spectroscopic
observations of the nearby elliptical galaxy NGC~821. The three data sets
have been shown to present a consistent picture of the stellar population 
within one effective radius. Strong  gradients in both age and metallicity 
were observed. The old ($\gtrsim$ 10 Gyr), low metallicity
([Fe/H]$\sim$--1.0) population in the outer 
regions of the galaxy (r$\sim 1r_{\rm e}$) exhibit no sign of the blue
horizontal branch morphologies present in some Galactic globular clusters of the
same age and metallicity. We have shown that
the large range of ages in the literature is probably the result of the
range of aperture sizes and the strong central age gradient.

We therefore find the central region of NGC~821 to have undergone a burst of
star formation between 1~Gyr and 4~Gyr ago. This conclusion is based on the
4~Gyr luminosity-weighted age found for the integrated spectra of the central
region and the observation that the galaxy, which shows no current star
formation, has also passed through the
post-starburst phase (which persists for approx. 1~Gyr). The old ages in the 
outer regions indicate that the burst was centrally concentrated and we have
shown that it can not have involved more than 10\% of the galaxy's stellar
mass.

We find the most likely scenario to be a very recent ($\sim$1~Gyr), low mass
($\sim$0.2\% of total galaxy mass) starburst in 
the in-situ gas, perhaps triggered by the accretion of a small (gas poor) satellite 
galaxy or a tidal interaction. In this scenario, the ambient gas in NGC~821 
was consumed in a centrally concentrated burst of star formation. The burst
would have had a high [Fe/H], but low [O/Fe], due to the long ($\gtrsim$10~Gyr) 
time-scale of star formation in the galaxy.\\

\noindent 

\noindent{\bf Acknowledgements}\\
These data were in part based on observations obtained at the Gemini
Observatory, which is operated by the Association of Universities for
Research in Astronomy Inc., under a cooperative agreement with the NSF on
behalf of the Gemini partnership: the National Science Foundation (United
States) the Particle Physics and Astronomy Research Council (United
Kingdom), the National Research Council (Canada), CONICYT (Chile), the
Australian Research Council (Australia), CNPq (Brazil) and CONICET
(Argentina). The Gemini program ID is 20041014-GN-2004B-Q-83.
The authors acknowledge the data analysis facilities provided by IRAF, which
is distributed by the National Optical Astronomy Observatories and
operated by AURA, Inc., under cooperative agreement with the National
Science Foundation. We thank the Australian Research council for funding 
that supported this work. K.G. acknowledges NSF CAREER grant
AST-0349095. We would like to thank Patricia
S\'{a}nchez--Bl\'{a}zquez for the provision of data and spectra.\\

\noindent{\bf References}\\
\noindent
Asplund M., Grevesse N., Sauval J., 2004, Cosmic abundances as records of 
stellar evolution and nucleosynthesis'', ed. F.N. Bash \& T.G Barnes, 
ASP Conference Proceedings, in press.\\
Bender R., Saglia R.P., Gerhard O.E., 1994, MNRAS, 269, 785\\
Bensby T., Feltzing S., Lundstr{\o}m I., 2003, A\&A, 410, 527\\
Bettoni D., Galletta G., García-Burillo S.,  2003, A\&A, 405, 5\\
Brown R.J.N.,  Forbes D.A., Kissler-Patig M., Brodie J.P., 2000, MNRAS, 317, 406\\
Caldwell N., Rose J.A., Concannon K.D., 2003, AJ, 125, 2891\\
Chiappini C., Romano D., Matteucci F., 2003, MNRAS, 339, 63\\
Denicol\`{o} et al. 2004, astro-ph/0411100 \\
Denicol\`{o} et al. 2005, astro-ph/0412435\\
de Souza R.E., Gadotti D.A., dos Anjos S., 2004, ApJS, 153, 411\\
de Vaucouleurs G., de Vaucouleurs A., Corwin H.G., Buta R.J., Paturel G.,
Fouqu\'{e} P., 1991, Third Reference Catalogue of Bright Galaxies,
Springer--Verlag, New York\\
Di Nella H., Garcia A.M., Garnier R., Paturel G., 1995, A\&AS, 113, 151\\
Emsellem E., Cappellari M., Peletier R.F., McDermid R.M., Bacon R., 
Bureau M., Copin Y., Davies R.L., Krajnovi D., Kuntschner H., Miller 
B.W., de Zeeuw, T.P., 1994, MNRAS, 352, 72\\
Forbes D.A., 1991, MNRAS, 249, 779\\
Forestall A., Gebhardt K., 2005, (in prep)\\
Fulbright J.P., Rich R.M., McWilliam A., 2004, Conference proceeding to
Nuclei in the Cosmos VIII, Vancouver\\
Georgakakis A., Hopkins A.M., Caulton A., Wiklind T., Terlevich A.I.,
Forbes D.A.,  2001, MNRAS, 326, 1431\\
Gonz\'{a}lez J.J., 1993, Ph.D Thesis, University of California, Santa Cruz\\
Gorgas J., Pedraz S., Guzman R., Cardiel N., Gonzalez J.J.,  1997, ApJ, 481, 19\\
Goudfrooij P., Hansen L., Jorgensen H.E., Norgaard-Nielsen H.U., de
Humphrey P.J., Buote D.A., 2005, ApJ, (submitted), astroph/0504008.\\
Jong T., van den Hoek L.B., 1994, A\&AS, 104, 179\\
J{\o}rgensen I., Franx M., Kj{\ae}rgaard P., 1995, MNRAS, 276, 1341\\
Greggio l., 1997, MNRAS, 285, 151\\
Kobayashi C., 2004, MNRAS, 347, 740\\
Mehlert D., Thomas D., Saglia R.P., Bender R., Wegner G., 2003, A\&A, 407, 423\\
Michard R., Prugniel P., 2004, A\&A, 423, 833 \\
Meza A., Navarro J.F., Steinmetz M., Eke V.R., 2003, ApJ, 590, 619\\
Proctor R.N., 2002, Ph.D Thesis, University of Central Lancashire, Preston, UK \\ 
  (http://www.star.uclan.ac.uk/~rnp/research.htm)\\
Proctor R.N., Sansom A.E., 2002, MNRAS, 333, 517\\
Proctor R.N,, Forbes D.A., Beasley M.A., 2004a, MNRAS, 355, 1327\\
Proctor R.N., Forbes, D.A., Hau G.K.T., Beasley, M.A., De Silva, G.M.,
  Contreras, R., Terlevich, A.I., 2004b, MNRAS, 349, 1381\\
Qian Y.-Z., Wasserburg G.J.,  2001, ApJ, 559, 925\\
Ravindranath S., Ho L.C., Peng C.Y., Filippenko A.V., Sargent W.L.W., 2001, AJ, 122, 653\\
Reda F.M., Forbes D.A., Beasley M.A., O'Sullivan E.J., Goudfrooij P., 2004, MNRAS, 354, 851\\
S\'{a}nchez--Bl\'{a}zquez P., Gorgas J., Cardiel N., Gonz\'{a}lez J.J.,
2005, MNRAS, submitted {\bf (SB05)}\\
Scorza C., Bender R., 1995, A\&A, 293, 20\\
Terlevich A.I., Forbes D.A., 2002, MNRAS, 330, 547\\
Thomas D., Maraston C., Bender R., 2003, MNRAS, 339, 897 {\bf (TMB03)}\\
Thomas D., Maraston C., Korn A., 2004, MNRAS, 351, L19  {\bf (TMK04)}\\
Tonry, John L.; Dressler, Alan; Blakeslee, John P.; Ajhar, Edward A.;
Fletcher A.B., Luppino G.A., Metzger M.R., Moore C.B., 2001, ApJ, 546, 681\\
Trager S.C., Worthey G., Faber S.M., Bustein D., Gonz\'{a}lez J.J., 1998, ApJS, 116, 1\\ 
Trager S.C., Worthey G., Faber S.M., Gonz\'{a}lez J.J., 2000, AJ 119, 1645\\
Vazdekis A., Peletier R.F., Beckman J.E., Casuso E., 1997, ApJS, 111, 203\\
Vazdekis A., Trujillo I., Yamada Y., 2004, ApJ, 601, L33\\
Worthey G., 1992, Ph.D. Thesis California Univ., Santa Cruz, USA\\
Worthey G., 1994, ApJS, 95, 107\\
Worthey G., Faber S.M., Gonz\'{a}lez J.J., 1992, ApJ, 398, 69\\

\label{lastpage}
\end{document}